\documentclass[twocolumn,showpacs,prl]{revtex4}
\usepackage{amsmath}
\usepackage{graphicx}
\usepackage{dcolumn}
\usepackage{bm}

\begin{document}

\title{Universal Description of Granular Metals at Low Temperatures: Granular Fermi
Liquid}
\author{I.~S.~Beloborodov,
A.~V.~Lopatin, and V.~M.~Vinokur}
\address{Materials Science Division, Argonne National
Laboratory, Argonne, Illinois 60439}

\date{\today}

\begin{abstract}
We present a unified description of the low temperature phase of
granular metals that reveals a striking generality of the low
temperature behaviors. Our model explains the universality of the
low-temperature conductivity that coincides exactly with that of
the homogeneously disordered systems and enables a straightforward
derivation of low temperature characteristics of disordered
conductors.

\end{abstract}

\maketitle

Granular metals exhibit a wealth of behaviors generic to strongly
interacting disordered electronic systems and offer a unique
experimental tool for studying the interplay between the effects
of disorder and interactions.  Depending on the strength of
coupling between the grains these systems can assume either
insulating- or metallic phases.  Remarkably, metallic samples
exhibit qualitatively different transport properties in different
temperature regimes; in particular, the low temperature phase
appears to be similar to disordered Fermi liquids.

The electronic transport in granular metals is governed by the
nontrivial interplay between the diffusive intra-grain electron
motion and grain-to-grain tunneling which is accompanied by
sequential charging of the grains involved in the particular
electron transfer process.  This brings the notion of the Coulomb
blockade, and one expects that it is the competition of
inter-grain coupling and electron-electron Coulomb interactions
that eventually controls transport properties of granular metals.
The basic parameter that characterizes transport properties is the
dimensionless tunneling conductance,  $g_{\scriptscriptstyle T}$.
Depending on the bare tunneling conductance $g_{\scriptscriptstyle
T}^{(0)}$, the conductivity can demonstrate either exponential
(insulating)-, at $g_{\scriptscriptstyle T}^{(0)} \ll
g_{\scriptscriptstyle T}^{\scriptscriptstyle C}$, or logarithmic
(metallic), at $g_{\scriptscriptstyle T}^{(0)} \gg
g_{\scriptscriptstyle T}^{\scriptscriptstyle C},$ temperature
dependencies~\cite{Beloborodov03,Efetov02,Gerber97,Simon87},
experiencing metal-insulator transition at $g_{\scriptscriptstyle
T}^{(0)} = g_{\scriptscriptstyle T}^{\scriptscriptstyle C}$.

The metallic phase was recently studied in
Ref.~\onlinecite{Beloborodov03} where it was shown that the low
temperature dependence of the conductivity of granular metals
coincides exactly with the corresponding result for the
conductivity of the homogeneously disordered samples.  A question
immediately arises: is it a coincidence that two different
physical systems exhibit identical low temperature transport
behaviors, or there is an underlying deep connection between the
two?  Furthermore, do all the other physical quantities (specific
heat, tunneling density of states, etc.) possess the same
universality?  The main result of our Letter is the answer to
these fundamental questions.

Generally speaking, all the universalities that one observes in
nature can usually be attributed to one or another kind of the
fundamental symmetry inherent to the physical system in question.
For example, all critical phenomena are described in terms of
universal models (Ginzburg - Landau Hamiltonian) that essentially
include only the information about the large scale symmetry of the
order parameter corresponding to the relevant degrees of freedom.
Thus in order to uncover the universality of inhomogeneous metals
one has to seek for a universal description in terms of the
appropriate large scale degrees of freedom that characterize
disordered conductors.

\begin{figure}[bp]
\resizebox{.43\textwidth}{!}{\includegraphics{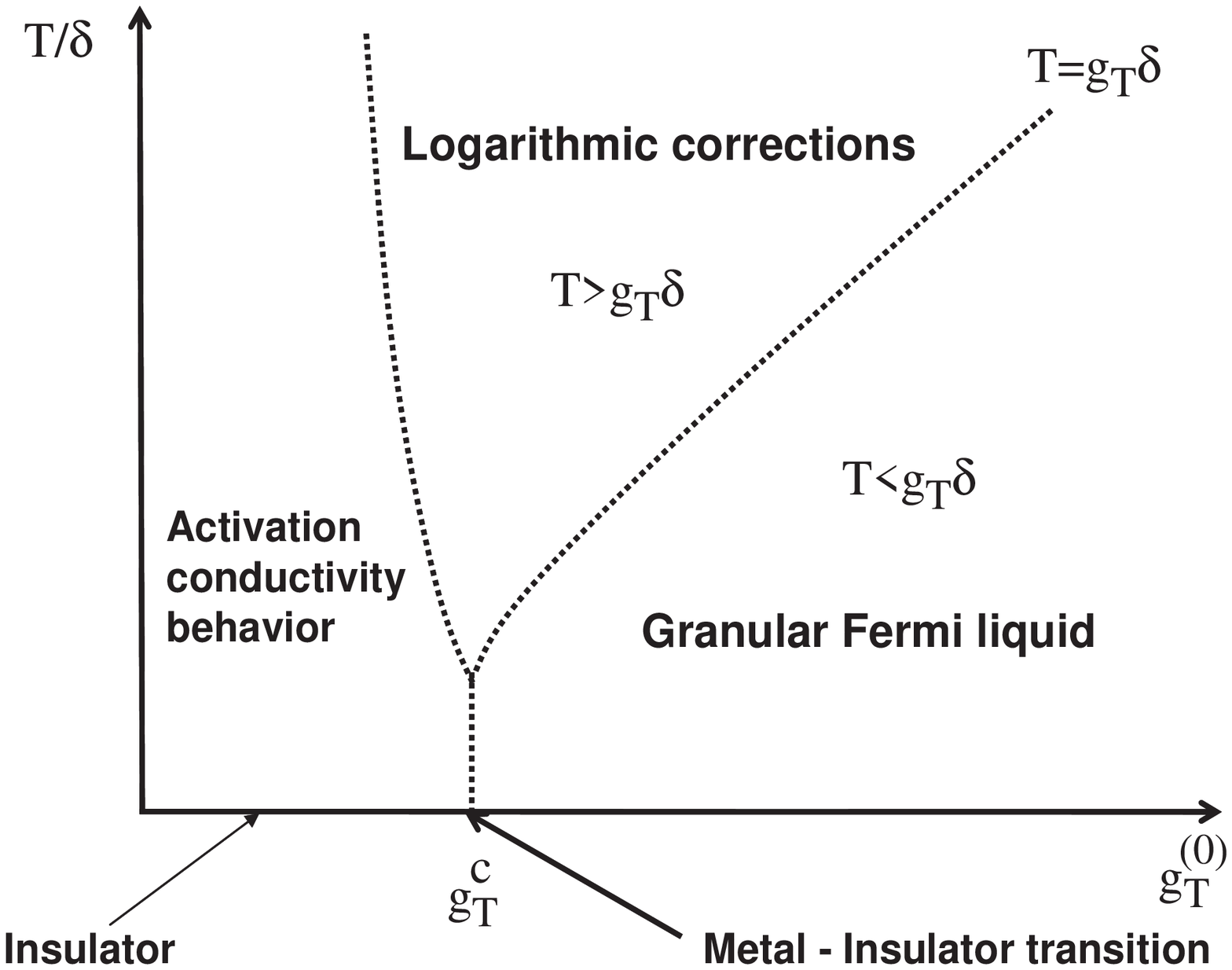}}
\vspace{0.6cm} \caption{Schematic phase diagram for granular
metals showing crossovers between three distinct phases. At small
bare tunneling conductances, $g_{\scriptscriptstyle T}^{(0)} <
g_{\scriptscriptstyle T}^{\scriptscriptstyle C}$, a granular metal
is in an insulating phase at zero temperature. The metal -
insulator transition in three dimensions at $T = 0$ occurs at
$g_{\scriptscriptstyle T}^{\scriptscriptstyle C} = (1/6\pi) \ln(
E_{\scriptscriptstyle C}/\delta)$, where $E_{\scriptscriptstyle
C}$ and $\delta$ are the charging energy and the mean energy level
spacing in a single grain respectively. At large tunneling
conductance, $g_{\scriptscriptstyle T}^{(0)}
> g_{\scriptscriptstyle T}^{\scriptscriptstyle C}$, the two different
types of conductivity behavior are possible: (i) the high
temperature phase, $T > g_{\scriptscriptstyle T}\delta$ is
characterized by the logarithmic temperature dependence of the
conductivity in all dimensions; (ii) the low temperatures phase,
$T < g_{\scriptscriptstyle T}\delta$ is the universal Granular
Fermi liquid phase described in terms of low energy interacting
diffusion modes.} \label{fig:2}
\end{figure}

We construct such a universal description of low temperature
physical properties of granular metals building on the
$\sigma$-model introduced first for the noninteracting dirty
metals in Refs.~\onlinecite{Wegner79,Efetov80} and generalized in
Ref.~\onlinecite{Finkelstein90} to include the interaction
effects. Our approach applies at temperatures $T
<g_{\scriptscriptstyle T}\delta$, where $\delta$ is the mean
energy level spacing in a single grain. The energy scale
$g_{\scriptscriptstyle T}\delta$ appears naturally as the upper
energy cutoff of the effective model, since
$\hbar/g_{\scriptscriptstyle T}\delta $ is the mean time for the
electron to escape from the granule~\cite{Beloborodov01}.

The main results of our work are as follows: Making use of the
effective description of the granular metals in terms of the
$\sigma$-model, Eq.~(\ref{L}), we show that: (i) All the phenomena
that are described in terms of the $\sigma$-model including
interaction and localization effects, and all the thermodynamic
quantities are universal for granular metals. (ii) There are
several physical quantities, which, although not directly related
to the charges in the $\sigma$-model can, nevertheless, be found
from the renormalization group equations describing the flow of
charges of the  $\sigma$-model. The important example of such a
quantity is the tunneling density of states. We summarized our
result in Fig.~\ref{fig:2} where three distinct phases that one
can identify on the basis of our approach are presented:  (i) The
universal low temperature phase, which we refer to as to the
granular Fermi liquid, generalizes naturally the Fermi liquid
phase of homogeneously disordered metals. The granular Fermi
liquid phase neighbors (ii) the high temperature, $T >
g_{\scriptscriptstyle T}\delta$, metallic phase governed by the
local single grain physics, and (iii) the insulating phase, where
$g_{\scriptscriptstyle T}^{(0)} < g_{\scriptscriptstyle
T}^{\scriptscriptstyle C}$, characterized by the activation
behavior of the conductivity.

Now we turn to the description of our model and  the derivation
the phase diagram in Fig.~\ref{fig:2}:  We consider a
$d-$dimensional array of metallic grains with the Coulomb
interaction between electrons. The motion of electrons inside the
grains is diffusive, and they can tunnel from grain to grain. We
assume that in the absence of the Coulomb interaction, the sample
would be a good metal. The system of weakly coupled metallic
grains is described by the Hamiltonian
\begin{subequations}
\label{hamiltonian0}
\begin{equation}
\hat{H} = \hat{H}_{0} + \hat{H}_{int} + \sum_{ij}
\,\hat{\psi}^{\dagger }({\bf r}_{i})\,t_{ij}\, \hat{\psi}({\bf
r}_{j}), \label{hamiltonian}
\end{equation}
where $t_{ij}$ is the tunneling matrix elements ($t_{ij}=t_{ji}$)
corresponding to the points of contact ${\bf r}_{i}$ and ${\bf
r}_{j}$ of $i$-th and $j-$th grains. The Hamiltonian $
\hat{H}_{0}$ in Eq.~(\ref{hamiltonian}) is
\begin{equation}
\hat H_0 = \sum \limits_i \int d^3 r_i
\,\hat{\psi}_{i}^\dagger\left[ \, \hat {\bf p}^2/2m - \mu + u({\bf
r}_i)\,\right] \hat{\psi}_{i}
\end{equation}
with $\mu$ being the chemical potential, it describes
non-interacting electrons scattered by random impurity potential
$u({\bf r}_i)$. The second term in the right hand side of
Eq.~(\ref{hamiltonian}) describes the electron-electron
interaction
\begin{equation}
\label{inetraction_term}
 \hat{H}_{int}={\frac{{\ e^{2}}}{{\
2}}}\,\sum_{ij}\,\hat{n}_{i}\,C_{ij}^{-1}\, \hat{n}_{j},
\end{equation}
\end{subequations}
where $C_{ij}$ is the capacitance matrix and ${\hat n}_i = \int
d^3 r_i \; {\hat \psi}_{i}^\dagger\, {\hat \psi}_{i} $ is the
electron number operator in the $i-$th grain.

Using Eqs.~(\ref{hamiltonian0}) the $\sigma-$model for granular
systems can be derived in a usual
way~\cite{Finkelstein90,Beloborodov01,Andreev03}: we decouple the
Coulomb interaction term in Eq.~(\ref{inetraction_term}) using the
axillary fields $V$, average over disorder introducing $Q-$matrix
field~\cite{Efetov80,Finkelstein90} and expand around the
diffusive saddle point. The final expression for the effective
low-energy action reads:
\begin{eqnarray}
\label{action} S &=& - \frac{\pi}{2\delta}\sum\limits_i {\rm
Tr}\left[ ({\hat \varepsilon} + V_i) Q_i \right]
\nonumber \\
&-&  {{\pi g_{\scriptscriptstyle T}}\over 8}\sum_{\langle
i,j\rangle }{\rm Tr}[ Q_i Q_j] + \frac{1}{T}\sum_{i,j}
V_i^*\frac{C_{ij}}{2e^2}\,V_j.
\end{eqnarray}
Here the sums are performed over the grain indices, the symbol
$\langle ... \rangle$ means summation over the nearest neighbors,
the trace is taken over spin, replica and Matsubara indices and
$\hat{\varepsilon} = i\partial_{\tau}$. The field $V$ in
Eq.~(\ref{action}) is a vector in the frequency and replica spaces
and the corresponding contraction is assumed: $V^*V =
\sum_{\omega_n,\alpha} V_{\omega_n,\alpha}^*\,
V_{\omega_n,\alpha}$, where $\alpha$ is the replica index and
$\omega_n = 2\pi T n$ are the bosonic Matsubara frequencies. The
$Q-$matrix in Eq.~(\ref{action}) is the matrix in the Matsubara,
spin, and replica spaces $ Q \to Q_{\omega_{n_1},\omega_{n_2};
\,\alpha, \beta }^{a,b} $ subject to constraint $Q^2=1.$ In
addition, each element of the $Q-$matrix is the quaternion, i.~e.
it can be presented as $ Q=q^i\,\tau_i$, where $q_i$ is the real
vector and $\tau_i$ are the quaternion matrices~\cite{Efetov80}.
For energies $\varepsilon \ll E_{\scriptscriptstyle T}$, where
$E_{\scriptscriptstyle T}$ is the Thouless energy, the
$Q-$matrices in Eq.~(\ref{action}) are coordinate independent
within each grain. In what follows we use Hikami-parametrization
for $Q-$matrix~\cite{Hikami}
\begin{equation}
Q=\left(  \begin{array} {cc} \sqrt{1 - B B^\dagger} & B \\
B^\dagger & -\sqrt{1-B^\dagger B }
\end{array} \right),
\end{equation}
where the matrix $B_{\omega_{n_1},\omega_{n_2}}$ has non-zero
elements only for frequencies $\omega_{n_1}
>0$, $\omega_{n_2} < 0$.
Expansion of the $Q-$matrix in powers of the field $B$ in
Eq.~(\ref{action}) provides a systematic way to take into account
$1/g_{\scriptscriptstyle T}$ corrections.

To derive the low energy $\sigma-$model that provides the
universal description of the low temperature phase, $T <
g_{\scriptscriptstyle T} \delta$, we apply the renormalization
group technique to the effective action in Eq.~(\ref{action}). We
divide the field $B$ into the slow and fast parts $B = B_s + B_f$
and define the fast part of the field
$B_{\omega_{n_1},\omega_{n_2}}$ in a way that it exists either for
$ \Lambda < \omega_{n_1}  < \Lambda + d \Lambda$ or $ -\Lambda - d
\Lambda < \omega_{n_2}  < -\Lambda.$ This procedure (the details
will be presented in the forthcoming publication) results in the
following renormalization group equation for the tunneling
conductance~\cite{Efetov02}
\begin{subequations}
\begin{equation}
\label{conductance_renorm}
 d g_{\scriptscriptstyle T}/d\ln \Lambda =  1/(2\pi d),
\end{equation}
where $d$ is the dimensionality of the granular array. Integration
of Eq.~(\ref{conductance_renorm}) in the energy interval
$(g_{\scriptscriptstyle T} \delta, E_C),$ where $E_C$ is the
charging energy for a single grain, results in the following
tunneling conductance
\begin{equation} \label{gC} g_{T} = g_{\scriptscriptstyle T}^{(0)} - (1/2\pi
d)\ln(E_{C}/\delta ).
\end{equation}
\end{subequations}
For energies lower than $g_{\scriptscriptstyle T} \delta$ the
physics is dominated by the distances that are much larger than
the size of a single grain. This allows us to consider the
continuum limit of the $\sigma-$model, and arrive at the
corresponding final expression for the action:
\begin{subequations}
\begin{eqnarray}
\label{L} S &=& - \frac{\pi}{2\delta} \int {\rm Tr}\left[
(\hat{\varepsilon} + V)Q - \frac{D}{4}(\nabla Q)^2
\right]\frac{d{\bf r}}{a^d} \nonumber \\ &+& \frac{1}{T}\int
\frac{d{\bf r}{d \bf r'}}{a^{2d}}{\rm Tr}\left[V^*_{\bf
r}\frac{C_{{\bf r}{\bf r'}}}{2e^2}V_{\bf r'} \right].
\end{eqnarray}
Here $a$ is the grain size and the trace is taken over the replica
indices. The renormalized diffusion coefficient $D$ in
Eq.~(\ref{L}) is given by
\begin{equation}
\label{coefD} D = g_{\scriptscriptstyle T}a^2\delta
\end{equation}
with $g_{\scriptscriptstyle T}$ being the renormalized tunneling
conductance from Eq.~(\ref{gC}).
\end{subequations}
Since the effective model~(\ref{L}) operates with the $Q-$matrices
which have only long-range degrees of freedom, it applies, with
the appropriate charges and upon the high-energy renormalization,
to any disordered metal, including a homogeneously disordered one.
Thus, all the information about the granularity of the sample is
hidden in the temperature independent renormalization of
coefficients of the effective model (\ref{L}). The conductivity of
the sample is related to the effective diffusion coefficient, $D$
through the usual Einstein relation
\begin{subequations}
\begin{equation}
\label{D} \sigma = 2e^2 D (a^d\delta)^{-1}.
\end{equation}
Effective model (\ref{L}) together with Eq.~(\ref{gC}) for the
renormalized conductance naturally explains the result for the low
temperature, $T < g_{\scriptscriptstyle T} \delta$, conductivity
obtained in Ref.~\onlinecite{Beloborodov03}.  The interaction
correction to conductivity has two contributions.  The first
contribution is temperature independent and is given by
\begin{equation}
\delta \sigma _{1} = -\sigma_0{\frac{{1}}{{\ 2\pi dg_{T}}}}\,\ln
\left[ {\frac{E_{C}}{\delta}}\right], \label{mainresult3}
\end{equation}
where $\sigma _{0}=2 e^{2}g_{T}a^{2-d}$ is the classical Drude
conductivity of granular metals. Equation~(\ref{mainresult3})
follows immediately from the renormalization of the tunneling
conductance $g_{\scriptscriptstyle T}$ in Eq.~(\ref{gC}) and is
specific to granular metals. The second contribution to
conductivity is temperature dependent and comes from the low
energy renormalization of the diffusion coefficient $D$ in the
effective model (\ref{L}). It coincides with the corresponding
correction to conductivity obtained for homogeneously disordered
metals in Ref.~\onlinecite{Altshuler}
\begin{equation}  \label{mainresult4}
\delta \sigma _{2}= \sigma _{0} \left\{
\begin{array}{lr}
{\frac{{\alpha }}{{12\pi ^{2}g_{T}}}}\sqrt{{\frac{{T}}{{g_{T}\delta }}}}
\hspace{1.6cm} d=3, &  \\
-\frac{1}{4\pi ^{2}g_{T}}\ln \frac{g_{T}\delta }{T}\hspace{1.4cm} d=2, &  \\
-{\frac{{\beta }}{{4\pi }}}\sqrt{{\frac{{\ \delta }}{{Tg_{T}}}}}
\hspace{ 1.9cm} d=1. &
\end{array}
\right.
\end{equation}
\end{subequations}
where $\alpha \approx 1.83$ and $\beta \approx 3.13$ are the
numerical constants.

Although the tunneling density of states is not directly related
to the charges in the $\sigma-$model, it can, nevertheless,  be
found from the renormalization group equations describing the flow
of charges of the $\sigma-$model. Since at low temperatures the
flow of coupling constants of the $\sigma$-model of granular
metals is determined by the same renormalization group equations
as in the case of homogeneously disordered metals, one arrives at
the important conclusion that the tunneling density of states has
a multiplicative structure:
\begin{equation}
\label{dos} \nu/\nu_0 = \nu_h \nu_l,
\end{equation}
where $\nu_0$ is the density of states of non-interacting
electrons, $\nu_h$ is the contribution to the density of sates
that comes from high energies, $\varepsilon >
g_{\scriptscriptstyle T}\delta$, while $\nu_l$ is the contribution
from low energies, $\varepsilon < g_{\scriptscriptstyle T}\delta$
which up to the proper renormalization of all constants coincides
with the corresponding result for the density of states of
disordered homogeneous metals. As an application of
Eq.~(\ref{dos}) let us consider the density of states of granular
films. The low energy contribution, $\nu_l$ in Eq.~(\ref{dos}), is
\begin{subequations}
\begin{equation}
\label{dos2} \nu_l = \exp\left[ -\frac{1}{16g_{\scriptscriptstyle
T}\pi^2}\ln \frac{g_{\scriptscriptstyle T}\delta}{T}\ln
\frac{g_{\scriptscriptstyle T}E_{\scriptscriptstyle
C}^4}{T\delta^3}\right],
\end{equation}
whereas the ``high energy" part of the tunneling density of
states, $\nu_h$ in Eq.~(\ref{dos}), is temperature independent at
$T < g_{\scriptscriptstyle T}\delta$ and for granular films is
given by
\begin{equation}
\label{dos3} \nu_h = \left[\frac{E_{\scriptscriptstyle
C}}{\delta}\right]^{1/\pi} \left[1 - \frac{\ln(E_C/\delta)}{4 \pi
g_{\scriptscriptstyle T}^{(0)}}\right]^{4g_{\scriptscriptstyle
T}^{(0)}}
\end{equation}
\end{subequations}
At large tunneling conductance, $g_{\scriptscriptstyle T} \gg 1$,
the result for the low energy contribution, $\nu_l$, in
Eq.~(\ref{dos2}) coincides with the perturbative result for the
density of states of granular metals obtained in
Ref.~\cite{Beloborodov03}.

Having described the effects of electron-electron interaction on
the transport properties of granular metals, we now turn to
quantum- (or weak localization) corrections to the
conductivity~\cite{Khmelnitskii79}. In the leading order in the
inverse tunneling conductance, $1/g_{\scriptscriptstyle T}$,
interaction and weak localization corrections can be considered
independently. As usual, the weak localization correction
$\delta\sigma_{\scriptscriptstyle WL}$ is defined by the following
expression
\begin{equation}
\label{wl23} \delta\sigma_{\scriptscriptstyle WL} = -
\frac{2}{\pi}e^2 g_{\scriptscriptstyle T}\int C(0,{\bf
q})\frac{d^d q}{(2\pi)^d},
\end{equation}
where $C(0,{\bf q})$ is the Cooperon propagator which in the
absence of electron-electron interaction is given by $C(\omega
,{\bf q}) = (D{\bf q}^2 - i\omega)^{-1}$ with $D$ being the
effective diffusion coefficient in Eq.~(\ref{coefD}). For $2D$ and
$1D$ samples it is important to take into account dephasing
effects since Eq.~(\ref{wl23}) diverges. Dephasing time
$\tau_{\phi}$ may be obtained from the effective model (\ref{L})
straightforwardly using the corresponding results for
homogeneously disordered metals~\cite{Altshuler82} with the proper
effective diffusion coefficient, $D = g_{\scriptscriptstyle T}
a^2\delta$. The final result for the weak localization corrections
reads
\begin{subequations}
\label{wl}
\begin{equation}
\label{wl1} \frac{\delta\sigma_{\scriptscriptstyle WL}}{\sigma_0}
= -\frac{1}{4\pi^2g_{\scriptscriptstyle
T}}\ln\left(\frac{g_{\scriptscriptstyle T}^2\delta}{T}\right),
\end{equation}
for granular films, and
\begin{equation}
\label{wl2} \frac{\delta\sigma_{\scriptscriptstyle WL}}{\sigma_0}
= -\frac{1}{2\pi g_{\scriptscriptstyle T}}
\left(\frac{g_{\scriptscriptstyle T}^2\delta}{T}\right)^{1/3}
\end{equation}
\end{subequations}
for granular wires. We notice that the quantum interference
corrections $\delta\sigma_{\scriptscriptstyle WL}$ may be easily
suppressed by applying relatively weak magnetic field such that
the main temperature dependence of the conductivity comes from
electron-electron interaction effects Eqs.~(\ref{mainresult3}) and
(\ref{mainresult4}).

In conclusion, we have derived the low energy  $\sigma$-model that
provides a universal description of the low temperature phase of
granular metals and, more generally, of any disordered conducting
medium. This model explains a striking similarity of the
low-temperature transport behaviors of different disordered
conductors as being governed by the same long wave electronic
diffusion modes. The proposed model enables one to derive the low
energy properties of granular metals from the corresponding
characteristics of disordered homogeneous metals. We demonstrated
the power of the developed approach by finding the density of
states, the interaction and localization corrections to the
conductivity of granular metals.

We thank K.~Efetov, L.~Glazman, D.~Khmel'nitskii and A.~Koshelev
for illuminating discussions. This work was supported by the U.~S.
Department of Energy, Office of Science through contract No.
W-31-109-ENG-38.

\end{document}